\PassOptionsToPackage{dvipsnames,table,xcdraw}{xcolor}
\documentclass[sigconf, nonacm]{acmart}

\acmYear{2026}

\usepackage{dirtytalk}
\usepackage[most]{tcolorbox}
\usepackage{soul} %
\usepackage{subfiles}
\usepackage{pifont} %
\usepackage{enumitem}
\usepackage{multirow}

\newcommand{\cmark}{\ding{51}}%
\newcommand{\xmark}{\ding{55}}%

\usepackage{xspace}

\newcommand{\AI}{AI\xspace}
\newcommand{\NoAI}{NoAI\xspace}
\soulregister{\AI}{0}
\soulregister{\NoAI}{0}

\usepackage{xcolor}

\definecolor{momentLtYellow}{RGB}{255,249,196} %
\definecolor{momentLtBlue}{RGB}{222,239,255}   %
\definecolor{momentLtGray}{RGB}{235,235,235}   %
\definecolor{momentOrange}{RGB}{217,95,2}      %
\definecolor{momentBlue}{RGB}{33,102,172}      %
\definecolor{momentGray}{gray}{0.35}           %

\newcommand{\momentBox}[3]{%
  \begingroup
  \setlength{\fboxsep}{2pt}%
  \colorbox{#1}{\textcolor{#2}{\sffamily\bfseries #3}}%
  \endgroup
}

\begin{document}

\newtcolorbox{finding}{
  enhanced,
  sharp corners,
  colback=white,             %
  colframe=gray,            %
  boxrule=0.5pt,
  left=3pt,
  right=3pt,
  top=3pt,
  bottom=3pt,
  before skip=3pt,
  after skip=3pt,
  fonttitle=\bfseries\small,
  coltitle=black,
  attach title to upper={},
  separator sign={\ },
}

\title{ChatGPT: Friend or Foe\\When Comprehending and Changing Unfamiliar Code}

\author{Norman Anderson}
\email{normananderson@uvic.ca}
\orcid{0009-0003-1238-8014}
\affiliation{
  \institution{University of Victoria}
  \city{Victoria}
  \country{Canada}
}
\author{Tarek Alakmeh}
\email{tarek.alakmeh@uzh.ch}
\orcid{0009-0008-5512-3549}
\affiliation{
  \institution{University of Zurich}
  \city{Zurich}
  \country{Switzerland}
}
\author{Victoria Jackson}
\email{v.jackson@soton.ac.uk}
\orcid{0000-0002-6326-931X}
\affiliation{
  \institution{University of Southampton}
  \city{Southampton}
  \country{UK}
}
\author{Guilherme Vaz Pereira}
\email{guilherme.v003@edu.pucrs.br}
\orcid{0009-0006-3521-6081}
\affiliation{
  \institution{Pontifícia Universidade Católica do Rio Grande do Sul}
  \city{Porto Alegre}
  \country{Brazil}
}
\author{Umit Akirmak}
\email{uakirmak@uvic.ca}
\orcid{0000-0003-3134-8133}
\affiliation{
  \institution{University of Victoria}
  \city{Victoria}
  \country{Canada}
}
\author{Anthony Estey}
\email{aestey@uvic.ca}
\orcid{0009-0002-0290-230X}
\affiliation{
  \institution{University of Victoria}
  \city{Victoria}
  \country{Canada}
}
\author{Rafael Prikladnicki}
\email{rafael.prikladnicki@pucrs.br}
\orcid{0000-0003-3351-4916}
\affiliation{
  \institution{Pontifícia Universidade Católica do Rio Grande do Sul}
  \city{Porto Alegre}
  \country{Brazil}
}
\author{Thomas Fritz}
\email{fritz@ifi.uzh.ch}
\orcid{0000-0002-1834-6240}
\affiliation{
  \institution{University of Zurich}
  \city{Zurich}
  \country{Switzerland}
}
\author{André van der Hoek}
\email{andre@ics.uci.edu}
\orcid{0000-0001-7917-932X}
\affiliation{
  \institution{University of California}
  \city{Irvine}
  \country{USA}
}
\author{Margaret-Anne Storey}
\email{mstorey@uvic.ca}
\orcid{0000-0003-2278-2536}
\affiliation{
  \institution{University of Victoria}
  \city{Victoria}
  \country{Canada}
}

\renewcommand{\shortauthors}{Anderson et al.}

\begin{abstract}
A rapidly growing body of research is examining how LLMs influence developers when they code. To date, this research has tended to focus on productivity and code quality outcomes, rather than the underlying cognitive processes involved in programming. 
To address this gap, we report on the results of an exploratory laboratory study of ten advanced student developers (five with support from AI and five without) who had to make a non-trivial extension to a sizable software system. Leveraging Polya's four problem-solving phases and 25 inductively-generated codes detailing distinct problem-solving behaviors as the primary lenses, we examined: (1) how AI impacted the problem-solving approach the developers used to solve the programming task, and (2) how AI impacted their progress when they became stuck.
For the analysis, we triangulated data across multiple sources (e.g., think-aloud, code changes, web searches, and LLM prompts).
Unexpectedly, while developers in the AI group repeatedly turned to the AI tool to offload certain aspects of the process, all detailed problem-solving behaviors appeared in both groups. 
We also found that nine out of ten participants found themselves stuck in their work, but with key differences in how they became stuck and unstuck. We highlight seven distinct causes for being stuck and highlight how AI in some cases helped and in other cases hindered becoming unstuck.

\end{abstract}

\begin{CCSXML}
<ccs2012>
   <concept>
       <concept_id>10011007.10011074</concept_id>
       <concept_desc>Software and its engineering~Software creation and management</concept_desc>
       <concept_significance>500</concept_significance>
       </concept>
 </ccs2012>
\end{CCSXML}

\ccsdesc[500]{Software and its engineering~Software creation and management}

\begin{CCSXML}
<ccs2012>
<concept>
<concept_id>10011007.10011074</concept_id>
<concept_desc>Software and its engineering~Software creation and management</concept_desc>
<concept_significance>500</concept_significance>
</concept>
</ccs2012>
\end{CCSXML}

\keywords{Code comprehension, Generative AI, LLM, problem-solving}

\maketitle

\section{Introduction}

Adding new features to an existing codebase is a complex task that requires a good understanding of the codebase to ensure code changes work as expected and do not lead to undesirable consequences. 
Yet the process of comprehending what is already there and how to best change it can be challenging and time-consuming,
in large part due to the vast amount of information and complex abstractions embedded in code \cite{Nam2024LlmCodeUnderstanding}, and the myriad ways in which a change can be made. 
To aid in what amounts to a problem-solving activity~\cite{koenemann1991expert},
developers adopt varying strategies~\cite{maalej_prog_comp_2014}, including steps such as extensive searching for pertinent information~\cite{ko2006exploratory}, detailing a plan~\cite{robillard_investigate_2004}, and making changes and testing them out~\cite{roehm_comprehend_2012}. It has already been noted that GenAI tools can help in this process, for instance, by explaining existing code \cite{Ebert2023GenAISoftwarePract}, providing relevant information on demand (e.g., details of an external API) \cite{Nam2024LlmCodeUnderstanding}, and suggesting code (e.g.,~\cite{mo_copilot_code_suggestions_2025}). How the use of GenAI potentially alters the problem-solving behaviors developers exhibit in addressing a change task, however, is little studied~\cite{qiao_slr_genai_code_comp_2025}.

We report on the results of an exploratory laboratory study with ten participants in which we compared the problem-solving behaviors employed by participants who had access to a GenAI tool (ChatGPT) to those who did not. We specifically sought to answer the following two research questions:

\begin{enumerate}
    \item[\textbf{RQ1}] 
    How does the use of \AI tools impact how developers problem solve in a complex programming task? 
    \item[\textbf{RQ2}] Why do developers get stuck and how can tools help or hinder them when they are stuck while making changes to unfamiliar code?
\end{enumerate}
The second research question recognizes that being stuck is a common occurrence in the problem-solving process and represents a particularly important moment since the process has broken down.

To answer our two research questions, we employed a multi-dimensional analysis centered on Polya's four problem-solving phases: 
(1) understanding the problem, (2) devising a plan, (3) carrying out the plan, and (4) looking back to evaluate the results (i.e., reflection) \cite{Polya1945HowToSolveIt}. Participants used the think-aloud method while they made a non-trivial change to an unfamiliar, relatively complex codebase. %
Moreover, participants' screens were recorded throughout the session and we conducted an exit interview during which we asked participants to make a system sketch and answer comprehension questions.
Using our various data sources, we used a mix of deductive and inductive coding to identify how each participant moved through Polya's phases, taking into consideration 25 specific observable behaviors exhibited by the participants as they went about their task with and without the AI.
We also coded when participants were stuck, what they attempted to do to become unstuck, and whether their attempts were successful.

Overall, considering RQ1 and Polya's problem-solving phases, we found participants with access to ChatGPT had a higher success rate and saved fewer but larger edits.
The use of \AI only led to a small number of differences in the observed 
problem-solving behaviors, and did not significantly impact the percentage of time the \AI group spent in any of Polya's phases.

Regarding RQ2 and stuckness, we found seven different causes of being stuck across Polya's phases and observe that the same (AI and non-AI) tools can both aid and hinder becoming unstuck, or even cause the stuckness in the first place, depending on the situation, use, and developer's approach. 

\section{Related Work}
\label{background}

\textit{Importance of Problem-Solving.}
Problem-solving is a core skill that influences performance and innovation across various domains.
Along with creativity and collaboration, %
it is essential for adapting to complex and dynamic environments \cite{GriffinCare2015ATC21S, DornerFunke2017ComplexPS, Mainzer2009Complexity21stCentury}. In the context of software engineering, Curtis et al. \cite{Curtis1986SoftwarePsychology} conceptualized programming as a complex cognitive activity that taps into how developers represent, understand, and solve problems. Although problem-solving is a core aspect of software engineering, it has not been systematically studied across different software development contexts \cite{Fagerholm2022CognitionSE}. Instead, most existing studies have focused on  cognitive abilities such as reasoning, memory, and cognitive load.
Previous work has also explored teaching problem-solving skills to novice programmers~\cite{loksa2016a, ko2019}.
Recently, attention has shifted toward performance outcomes, especially within the context of human-AI collaboration. For example, both Boussioux et al. \cite{BoussiouxLéonard2024TCFG} and Dell’Acqua et al. \cite{dellacqua2023navigating} examined the effects of generative AI on the resulting productivity and quality in professional tasks. Boussioux et al. \cite{BoussiouxLéonard2024TCFG} showed that AI assistance increased productivity and creativity in problem-solving. Dell’Acqua et al. \cite{dellacqua2023navigating} similarly found that AI improved performance and the quality of work on familiar or well-defined tasks, though it was less effective in more complex or unfamiliar tasks.

\textit{Theoretical Foundations of Problem-Solving.}
Problem-solving is characterized
as cognitive searching within a problem space, where individuals move from an initial state to a goal state through the application of strategies and heuristics \cite{NewellSimon1972HumanProblemSolving}.
It is broadly understood as a goal-directed process involving cognitive and affective components shaped by both task characteristics and a solver’s prior knowledge, motivation, and metacognitive regulation~\cite{Davidson1994, HeppnerKrauskopf1987PersonalProblemSolving,  Schoenfeld1992LearningToThink}.

Polya’s \textit{How to Solve It} \cite{Polya1945HowToSolveIt} translated the theoretical study of problem-solving into a practical heuristic framework. 
It provides a lens for examining how individuals plan, implement, and refine solutions in digital problem-solving environments, including software engineering \cite{Wu2024ProblemSolvingCT}.
Understanding the problem involves clarifying the task requirements and identifying the variables, inputs, and outputs \cite{PENNINGTON1987295, 402076} to conceptually understand `how the system works' at varying levels of abstraction~\cite{HEINONEN2023107300}. Devising a plan involves forming hypotheses about program behavior and planning the steps needed to reach a solution \cite{PENNINGTON1987295},
and revising or replacing plans which do not lead to desired outcomes. Effective execution of a plan depends on the problem solver's technical skills, cognitive focus, and emotional patience \cite{DornerFunke2017ComplexPS, funkeComplexProblemSolving2010}.

\textit{Human–AI Collaboration and Emerging Perspectives.}
GenAI tools can enhance access to information and support completion of complex tasks~\cite{PaulVariawa2025GenerativeAIFramework}, but may also reduce cognitive engagement or critical evaluation with over-reliance~\cite{Hou2025GenerativeAIScale, Kasneci2023ChatGPTforGood}. Neurocognitive evidence suggests reduced activation in control networks during AI assistance, consistent with ``cognitive debt''~\cite{Kosmyna2025YourBrainOnChatGPT}.
Interactions with GenAI tools range from reflective and cautious all the way to overly dependent reliance~\cite{Hou2025GenerativeAIScale}, and the impact of this collaboration depends on the types and number of interactions. Most work exploring human-AI collaboration emphasizes attitudes or outcomes rather than process-level cognition. Moreover, a recent literature review~\cite{qiao_slr_genai_code_comp_2025} on the use of GenAI for supporting code comprehension identified that much code comprehension research focused on GenAI support for explaining software. We differentiate our work by asking how AI reshapes the sequence, depth, and regulation of problem-solving activities in realistic software contexts.

\textit{Flow, Stuck, and Self-Regulation.}
GenAI changes both \textit{when} and \textit{how} developers get stuck. While it can offer suggestions that appear to unblock progress, it may also shift the impasse from implementation to comprehension or verification. Studies found that AI-assisted programming accelerates task completion overall, but also leads novices toward false progress, illusion of competence, or premature acceptance of correctness~\cite{10.1145/3617367, 10.1145/3632620.3671116, 10.1145/3491101.3519665},
whereas  experienced users engage more selectively and verify more effectively~\cite{10.1145/3754508.3754512, 10.1145/3750069.3750397}.
From this perspective, the quality of human–AI collaboration depends not only on correctness or efficiency, but on how interactions regulate transitions into and out of stuck states.
Analyzing those transitions offers a process-level lens that complements outcome metrics of performance or comprehension~\cite{Kosmyna2025YourBrainOnChatGPT, 7194617, 10.1145/3632620.3671116, 10.1145/3617367}. 
Consistent with classic views of impasse and restructuring~\cite{Duncker1945ProblemSolving, Polya1945HowToSolveIt}, we treat stuckness as a trigger for representational change, and our characterization specifies
how such impasses arise and are resolved within software engineering work with and without AI assistance.

\section{Methodology}
\label{methodology}

Our laboratory-based experiment used a between-subjects design with two conditions:
(1) \textbf{AI condition:} participants could use ChatGPT and the web (AI group); and (2) \textbf{No-AI condition:} participants could use the web but not ChatGPT (\NoAI group).  %
The participants were required to make a change across a complex codebase to be representative of the type of real problems faced by developers in the workplace.
The study received institutional ethics approval, with participants providing written informed consent and all recordings and data stored on secure institutional servers accessible only to the research team.
All non-participant-identifying research data, including task description, source code, participant component sketches, anonymized artifacts, and ChatGPT logs, are publicly available~\cite{supp_data}.

\subsection{Task Design}
\label{study-design}

Our experiment %
required participants to make a non-trivial change to a realistic and functional codebase specifically developed for our study. Modeled after an existing open source enterprise-grade system~\cite{blazity2024}, the codebase is a TypeScript monorepo comprising a Next.js and React frontend, an Express backend, and an SQLite database. The codebase consists of 35k lines of TypeScript code across 227 files.
We included a couple of partially implemented features to mimic the in-progress state of a junior web developer's feature branch. We also provided a README file to emulate instructions from a team lead, including background context, task acceptance criteria, and links to relevant source files. 

The specific change task was to fix an issue with user registration.
Successful completion of the task required running the application locally, registering a new user through the interface, and verifying the expected confirmation message---initially absent because of an error message identifying deprecated API usage, thus requiring further changes. Addressing the full change task involved creating three new functions, modifying one existing function, and modifying the application's routes, across five distinct files throughout the codebase. Moreover, it required participants to work with three external libraries and understand the purpose of the two existing TypeScript interfaces defined to represent user accounts. 

\subsection{Participants}
\label{participants}

We recruited eleven participants (identified as P1–P11 throughout the paper) from a large public university in Canada and through the researchers’ networks. One participant (P8) withdrew during the session, leaving ten for final analysis. Each participant received compensation equivalent to CAD 35.
All participants were advanced students with industry experience as junior developers, reporting an average of 2.56 years of industry programming experience (range: 0.6–8 years), and communicating primarily in English for their job and/or studies. 
Participants were randomly assigned to the \AI and \NoAI conditions, resulting in a balanced final sample (five participants per group).
Table~\ref{tab:participants} summarizes participants’ demographics, including industry and personal programming experience.

\begin{table}[ht!]
    \centering   
    \caption{Demographics of the ten participants.}
    \begin{tabular}{cccc}    
    \toprule
        \textbf{ID} & 
        \textbf{Gender} & 
        \begin{tabular}[c]{@{}c@{}}\textbf{Industry}\\\textbf{Exp (years)}\end{tabular} &
        \begin{tabular}[c]{@{}c@{}}\textbf{Non-Industry}\\\textbf{Exp (years)}\end{tabular}
        \\
    \midrule
       P1  & Male  & 1 & 4\\
       P2  & Male  & 8 & 10\\
       P3  & Female & 1 & 4\\
       P4  & Male  & 1.25 & 0\\
       P5  & Female & 2 & 7.5\\
       P6  & Female & 0.66 & 3.5\\
       P7  & Unanswered  & 5 & 10\\
       P9  & Male & 1.83 & 5\\
       P10  & Male & 4 & 4\\
       P11  & Female & 1 & 3\\
    \bottomrule
    \end{tabular}
    \label{tab:participants}
\end{table}

\subsection{Procedure}

\label{procedure}
The study was conducted between February 5 and May 1, 2025. All sessions were held in person and facilitated by the same researcher to ensure protocol consistency. Before the main study, we conducted four pilot sessions to validate the task setup, timing, and data collection instruments, leading to minor adjustments. 

Each study session followed a three-phase structure: (1) orientation, (2) task execution, and (3) post-task questionnaire and system sketch. 
Before the scheduled session, each participant completed a remote pre-survey designed to collect demographic information and assess their familiarity with the technologies used in the study. 
During the \textit{orientation}, participants received a short walkthrough of the repository structure, project context, and local execution instructions. They were informed about the study procedure and reminded that they could take breaks at any time. A pen and paper were given to each participant for optional note-taking during the session. During \textit{task execution}, participants performed the assigned programming task locally in VS Code while following a think-aloud protocol \cite{ericsson2011thinking}.  Those in the \AI group were provided access to a ChatGPT Plus account with the default model (gpt-4o) selected and unrestricted use of available features. However, to avoid interference, all other AI autocomplete or assistant tools (e.g., GitHub Copilot) were disabled. Those in the \NoAI group were not permitted to access any AI assistant tools, including AI overviews in web searches.
On two occasions, P4 mistakenly accessed an AI search overview, but we did not disqualify their participation because they only cursorily reviewed both overviews and appeared to find them unhelpful.
The facilitator remained silent and out of the participant’s view throughout the task, taking notes. 
Sessions were timed at three hours, with a short break offered around the midpoint (approximately 60 minutes into the task execution). While some participants took the break, others chose to continue working. 
Finally, when 20 minutes remained in the session or after completing their task, participants answered a \textit{post-task questionnaire} assessing difficulty, frustrations encountered, and confidence. They also answered predefined comprehension questions to assess their understanding of the codebase and task requirements. In addition, each participant was asked to sketch the high-level architecture of the system to externalize their mental model of how the components interacted.

\subsection{Data Analysis}
\label{analysis}

Some of our analysis informed both research questions. We present this before detailing the core analyses we did for each question.
We summarize the data we collected in Table~\ref{tab:datacollection} and how the different types of data helped us answer our research questions.

\begin{table*}[t]
\centering
\caption{Data collected and corresponding analytical focus for our research questions.}
\label{tab:datacollection}
\rowcolors{2}{gray!15}{white}
\begin{tabular}{p{3cm} p{6.5cm} p{6.5cm}}
\toprule
\textbf{Data Type} & \textbf{Description} & \textbf{Used to Examine} \\
\midrule
Screen + audio recordings & Full Zoom recordings of participant interaction with the codebase and tools & Problem-solving and comprehension behaviors (RQ1), evidence of being stuck (RQ2) \\
ChatGPT transcripts (AI condition only) & Timestamped prompt--response logs from the provided account & Help-seeking patterns, evidence of being stuck, AI influence on reasoning (RQ2) \\
Source code diffs & Local Git changes captured in terms of LOC changed, and intervals between changes  & Technical progress, solution paths (RQ1) \\
Post-task questionnaires & Likert-scale items and open-text responses on experience and confidence &  Code comprehension (RQ1), self-reported difficulties, ways of getting unstuck (RQ2) \\
Architecture sketches & Scanned participant diagrams of task-relevant architecture & Code comprehension and understanding (RQ1, RQ2) \\
Field notes & Facilitator observations recorded during the sessions & Interaction context, notable behaviors, qualitative insights (RQ1, RQ2) \\
\bottomrule
\end{tabular}
\end{table*}

\subsubsection{Setting the Stage: Descriptive Context for RQ1 and RQ2}
\label{comp-description}

To provide context for our two research questions, we collected descriptive data to describe \textbf{task success} and \textbf{completion times}.
We analyzed the task's VS Code file history logs of each group (\AI and \NoAI\kern-2pt) to determine characteristics of their code changes (i.e., number of saved changes and time interval between them).
Additionally, participants were assigned a \textbf{code comprehension} score based on their post-task questionnaire responses and architectural sketches. We evaluated responses across two dimensions: \textbf{system understanding}, the ability to identify high-level component responsibilities, and \textbf{technical fluency}, the depth and accuracy of implementation details in their responses (i.e., naming specific functions or identifying precise points of failure). Two researchers (independently and in agreement) assigned comprehension scores as follows:
\begin{itemize}[label=-, nosep, leftmargin=*]
    \item \textbf{High}: Strong system understanding and technical fluency, characterized by detailed and accurate call-stack reconstructions.
    \item \textbf{Med}: A gap in either system understanding or technical fluency (i.e., accurate high-level idea but limited implementation detail).
    \item \textbf{Low}: Foundational gaps in both system understanding and technical fluency, resulting in incorrect or incomplete responses.
\end{itemize}

\subsubsection{Analysis for Identifying Problem-Solving and Comprehension Behaviors (RQ1) }

To identify when participants were in a particular \textbf{phase} of the problem-solving process as defined by Polya \cite{Polya1945HowToSolveIt}, we first developed a fine-grained set of codes and then clustered these into the four phases.
The fine-grained codes were inductively formed by analyzing three sessions, and subsequently applied deductively to the remaining seven sessions. 
One researcher coded two participants inductively (P9 and P10), identifying a set of initial codes, iterating over two more passes, and producing a refined set of codes (e.g., interpreting, recognizing, hypothesizing, deciding, solving, reusing) that are indicative of which Polya phase a participant may be in. 
The researcher then clustered these initial fine-grained codes into the phases of the Polya problem-solving model. After,
two researchers jointly walked through P9 \textit{again} as well as another participant (P2) to review and refine the codebook.  The two researchers then independently recoded P9 and P2, and then realigned the codes based on the disagreements in the  fine-grained, nuanced codes. Some minor differences were also noted about when start/end times were recorded for different behaviors. These differences could influence the interpretation of timing-related results, however, our focus remained on examining relative patterns across participants rather than precise durations. The final version of the resulting codebook is shown later in the paper in Table~\ref{table-codebook}. 

Our qualitative coding of \textbf{observable development behaviors} (such as searching, hypothesizing, deciding) for the remaining eight participants was then performed deductively using the fine-grained codes (coding four each independently),
recording times of the start and end of each phase for detailed and Polya codes.  At the same time, while coding, extensive \textbf{memos} were noted down (e.g., when we noted participants had difficulties, capturing details of how they recovered, and how the tools they used either helped or hindered their progress). %
We then cross-checked this behavior by analyzing the web searches all participants made and the prompts issued to ChatGPT by those in the \AI group, again making extensive memos that serve as the basis for the observations we share in Section 4.

\subsubsection{Analysis of Being Stuck (RQ2)}
To answer RQ2, we used the: (1) coded screen recorded sessions with think-aloud utterances, (2) ChatGPT chat logs for the \AI participants, and (3) extensive memos recorded during qualitative coding from when we noted participants had difficulties. Our memos (in combination with the coding of problem-solving phases and detailed comprehension behaviors) led to the insight and identification of ``stuck moments'' (reasons for getting stuck) and how the tools they were using helped or hindered their progress. These stuck moments emerged inductively and iteratively as we conducted our detailed analysis, which consisted of two researchers jointly working through the memos, revisiting sessions to find additional occurrences of participants being stuck, and developing a typology of the stuck moments together with the roles the tools played in sometimes leading them to get stuck, sometimes helping them get unstuck, and sometimes causing them to be even more stuck. 
The post-task questionnaire also provided us with additional insights on when developers felt they were stuck and how they became more or less stuck.

\section{Findings}
\label{findings}

In this section, we answer our two research questions based on the results from performing the analyses described in the prior section.

\subsection{RQ1: AI and NoAI Problem-Solving}
\label{rq1}

To answer our first research question, \textit{``How does the use of \AI tools impact how developers problem solve in a complex task''}, 
we first report some descriptive data 
on the nature of the code changes made, the participants' success rate, and their level of code comprehension.  
We then detail how they approached the change task, first through the lens of Polya's four high-level problem-solving phases, and then through the detailed 25 observable development behaviors.

\subsubsection{Participant Success and Comprehension}
\label{descriptive-statistics}

Every participant dedicated time at the beginning of their task to read the initial instructions. Each was told they could complete the task in any order; 9/10 participants chose to address the requested changes linearly. Whenever participants saved a file, VS Code recorded a copy of the file's contents. 
Our analysis of these file versions saved by VS Code revealed developers with \AI support saved their files less than half the number of times compared to the \NoAI group (134 vs. 290), with the average number of lines added and/or removed per save greater for the \AI group (14.9 LOC) than the \NoAI group (5.7 LOC). Furthermore, the AI group tended to have longer intervals between saved changes (average of 2.65 minutes, median of 1.35 minutes) compared to the \NoAI group (average of 1.43 minutes, median of 0.62 minutes).

Four of the ten participants successfully completed the task (three 
in the \AI condition, and one in the \NoAI condition), correctly making the necessary changes within the five files that were part of the codebase's incomplete registration feature. 
As shown in~\autoref{tab:summary_task_completion}, the times to complete the task varied, as did the number of prompts entered by the participants in the \AI condition. 
Even for the participants who did not finish, times were different, as they elected to `give up' at different moments (e.g., P6).

\begin{table}
\caption{Summary of task completion and assessed code comprehension for all participants. The first five shown are non-AI, the next five are AI.}
    \centering
    \begin{tabular}{p{0.05\linewidth}p{0.09\linewidth}p{0.14\linewidth}p{0.18\linewidth}p{0.11\linewidth}p{0.1\linewidth}}
    \toprule
    \textbf{PID} & \textbf{Success} & \textbf{\#Searches/ Prompts} & 
    \textbf{Task Time (min:sec)} & 
    \textbf{\#File Saves} &
    \textbf{Code Comp.}\\
    \midrule
    P1 & \xmark & 28 & 120:59 & 59 & Med\\
    P4 & \xmark & 24 & 131:21 & 95 & Med\\
    P5 & \xmark & 36 &138:41 & 43 & Med\\
    P6 & \xmark & 6 & 74:07 & 59 & Low \\
    P9 & \cmark & 2 & 29:16 & 34 & High\\
    \midrule
    P2 & \cmark & 11 & 47:56 & 27 & High\\
    P3 & \xmark & 29 &117:17 & 17 & Low\\
    P7 & \cmark & 28 &137:15 & 34 & Med\\
    P10 & \cmark & 38 &75:23 & 14 & Med\\
    P11 & \xmark & 16 & 121:04 & 42 & Med\\
    \bottomrule
    \end{tabular}
    \label{tab:summary_task_completion}
\end{table}

\begin{figure}[b]  
  \includegraphics[scale=0.35]{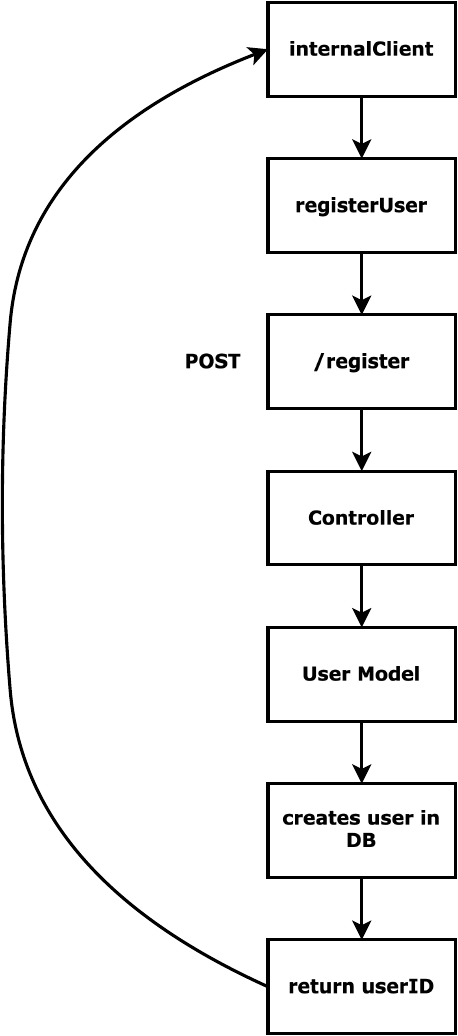}\hfill
  \includegraphics[scale=0.45]{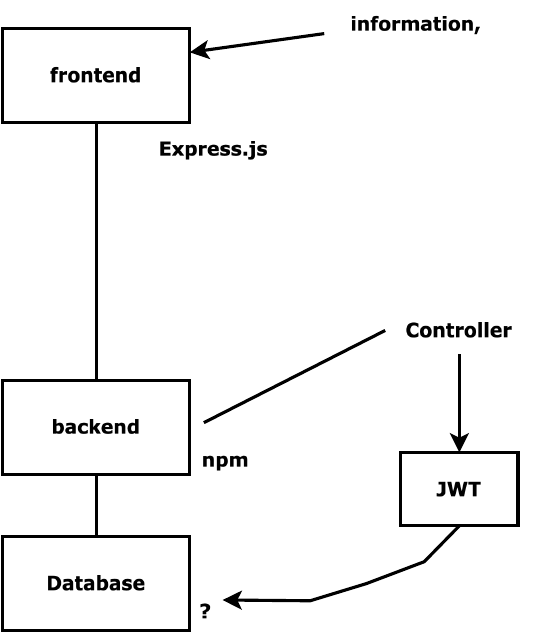}
  \caption[Component sketches from P2 and P3]{Component sketches from P2 (left) and P3 (right), redrawn to protect participants' anonymity.}
  \Description[Component sketches from P2 and P2]{P2 has a detailed component sketch showing many different components involved when trying to register a user from the front end. Shows the internal client, register user, the route, the controller, the user model and inserting int the database. In comparison, P3 is more sparse showing only a frontend, backend and database with a few notes about a controller and JWT (not relevant to the task).}
  \label{fig:p2-p3-system-diagrams}
\end{figure}

As the task concluded, participants exhibited varying levels of understanding on the code comprehension assessment (see Sec.~\ref{comp-description}).
P2 had a high code comprehension score; their sketch shows all key components and their relationships. Compare their sketch (\autoref{fig:p2-p3-system-diagrams}, left) to P3's (right), who had a low comprehension score, with many missing components and some relationships incorrect.

Successfully completing the task did not always mean the participant had a reasonable (i.e., high or medium) level of system understanding, nor did a reasonable level of system understanding always lead to successful completion.  As an example of the former, P9 (NoAI - high code comprehension) had a more complete architecture diagram than P10 (medium), who missed a few key components (AI - med code comprehension), but both successfully completed the task. 
As an example of the latter, P11 (AI) also had a medium level of system understanding like P10, but failed to complete the task. Interestingly, all of the participants with a high understanding completed the task, while all with a low understanding failed, regardless of \AI or NoAI.  For the medium level of understanding, all participants in the \NoAI group failed, whereas two out of three in the \AI group succeeded. A common factor among participants who failed was that they appeared to struggle with understanding the technology stack and/or exhibited a lack of experience with the programming language--it appears that AI may have helped some of those with a medium understanding overcome this deficiency.

\begin{figure*}[t]
  \centering
  \includegraphics[width=1\textwidth]
  {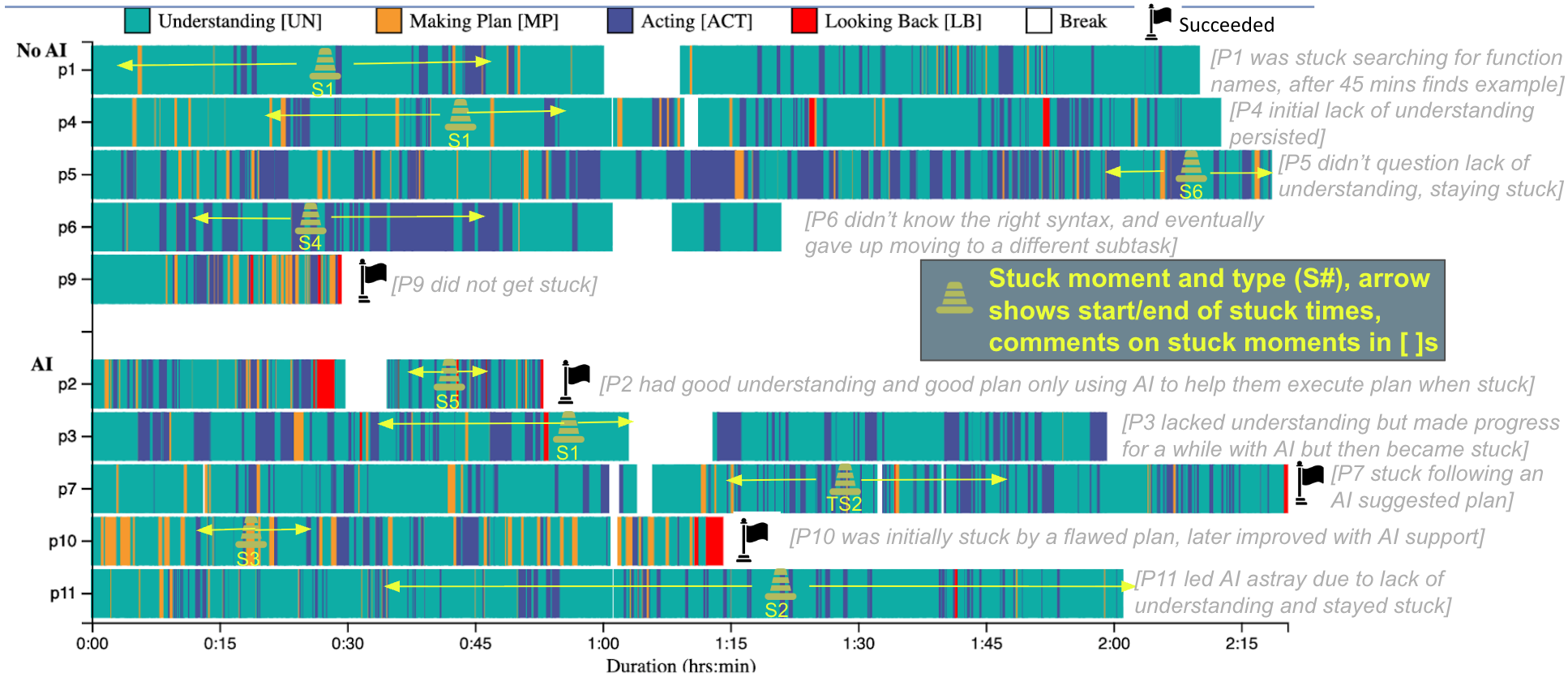}

  \caption{Timeline summarizing the time each participant spent in and across the four Polya stages throughout their task, whether they succeeded, and one example of when and how participants became ``stuck'' (discussed in Section~\ref{rq2}).}
  \Description[Polya state timeline]{The multicolored state timeline shown groups NoAI and AI participants separately, with four colors representing the four Polya stages of Understanding [UN], Making Plan [MP], Acting [ACT], and Looking Back [LB]. It is observed that NoAI participants spend relatively more time in the Acting stage, in comparison to AI participants.}
  \label{fig:polya_state_timeline}
\end{figure*}

\begin{finding}
{\textbf{Findings:}}
3 of 5 participants in the \AI group completed the task, as did 1 of 5 in the \NoAI group. Notably, 2 of the successful participants who used \AI only had a medium comprehension score, yet all 3 \NoAI participants with a medium score failed to complete the task. Also, the \NoAI participants made twice as many file saves, with much smaller edits, than the \AI group.
\end{finding}

\subsubsection{Polya's Phases} %
\label{sec:polya_phases}

Above, we reported the success rate of the participants at the assigned task, their number of file saves (which differs across the \AI and \NoAI conditions), and their assigned comprehension scores.  However, with just ten participants, it is not possible to conclusively attribute the increased success rate to the use of AI.  
Here, we discuss how we used Polya's problem-solving phases to illustrate how they solved the tasks.
 \autoref{fig:polya_state_timeline} shows when each participant spent time in each of the four Polya phases. Both groups (\AI and \NoAI) spent the majority of their time\footnote{The reported values represent the \textit{proportion} of each participant's total task time spent in each phase, rather than raw durations.} in the \textbf{understanding} phase  (\NoAI: M=0.67, SD=0.13; \AI: M=0.69, SD=0.10), followed by the \textbf{acting} phase (\NoAI: M=0.25, SD=0.13; \AI: M=0.22, SD=0.08).  The time spent in the \textbf{planning} and \textbf{looking back} phases was very low for both groups, with each phase averaging below 0.08. This pattern suggests that participants strongly focused on understanding and action rather than on planning and reflection. As several participants completed only a portion of the task, their opportunity for reflection was naturally lower, since not all sessions reached a partial or completed solution to reflect on. 

Participants in both groups spent on average a quarter of their time in the acting phase. Since more participants in the \AI group finished (or almost finished) the task than in the \NoAI group, this may indicate that their use of ChatGPT sped up progress within the acting phase. This matches our observations that the \NoAI group engaged more in trial-and-error programming, as they saved their changes more frequently (see Sec. 4.1.1), whereas some participants in the \AI group were able to generate larger blocks of code at a time that they only needed to review and update (see Sec. \ref{sec:dev_activities}).

\begin{finding}
{\textbf{Findings:}}
Developers (across both groups) spent the majority of their time in the understanding phase, followed by time in the acting phase, with much less time in the planning and looking back phases. 
Given both groups spent similar percentages of time in each phase, the \AI group's higher completion rate may be attributed to external guidance they received from ChatGPT. 

\end{finding}

\subsubsection{Observed Behaviors}
\label{sec:dev_activities}

\begin{table}[t]
\caption{Codebook of inductively derived codes, or \textit{observed behaviors}, per Polya phase. Column 3 indicates the number of participants in the \AI group for which the behavior was observed, column 4 does the same for the NoAI group. GPTX, AIP, REFA, and DEL are applicable to the \AI group only.}
\label{table-codebook}
\footnotesize
\renewcommand{\arraystretch}{1.05}
\begin{tabular}{p{2.0cm} p{3.5cm} cc}
\toprule
\textbf{Polya Phase} & \textbf{Inductive Code} & \textbf{AI} & \textbf{NoAI} \\
\midrule

\multirow{10}{=}{Understanding (UN)}
 & ASK: Asking question                        & 5 & 5 \\
 & BOT: Bottom-up comprehension        & 5 & 5 \\
 & INT: Interpreting instructions                  & 5 & 5 \\
 & SEA: Searching                      & 5 & 5 \\
 & HYP: Hypothesizing                  & 5 & 4 \\
 & AHA: Aha moment                     & 4 & 4 \\
 & DEB1: Reading logs/errors           & 4 & 4 \\
 & WEBX: Web helps understand & 2 & 5 \\
 & TOP: Top-down comprehension         & 1 & 3 \\
 & REK: Recognizing technology                     & 2 & 1 \\
 \cmidrule{2-4}
 & GPTX: ChatGPT helps understand & 4 & n/a \\
\midrule

\multirow{5}{=}{Making plan (MP)}
 & DCO: Decomposing problem                  & 5 & 5 \\
 & DEC: Deciding                      & 5 & 5 \\
 & ASU: Making assumption                    & 3 & 4 \\
 & OUT: Outlining steps               & 1 & 5 \\
 \cmidrule{2-4}
 & AIP: Delegating planning to AI                   & 2 & n/a \\
\midrule

\multirow{6}{=}{Acting (ACT)}
 & REU: Reusing from codebase                         & 5 & 5 \\
 & SOL: Writing code to solve                       & 5 & 5 \\
 & DEB3: Inserting debug statement                    & 4 & 3 \\
 & RUN3: Running partial solution                   & 4 & 2 \\
 & REFW: Referencing web           & 1 & 5 \\
 \cmidrule{2-4}
 & REFA: Referencing \AI suggestion           & 5 & n/a \\
 & DEL: Delegating coding to AI       & 3 & n/a \\
\midrule

\multirow{2}{=}{Looking back (LB)}
 & RUN4: Running complete solution                & 5 & 1 \\
 & REFL: Reflecting                  & 1 & 3 \\
\bottomrule
\end{tabular}
\end{table}

The groups shared a majority of observed behaviors (\autoref{table-codebook}). Excluding \AI-specific observable behaviors, all five participants in each group performed 8 of 21 behaviors, and a further five behaviors were performed by at least 3 out of 5 in each group. The three most unbalanced behaviors between the two groups were: (1) GPTX together with WEBX, indicating the AI group leaned primarily on ChatGPT to help them understand, even though a couple of \AI participants also used the web; (2) OUT, which all \NoAI participants performed explicitly by outlining their plan for a change in comments or otherwise, versus only one of the \AI participants who did this themselves, two other AI participants delegating it to ChatGPT, and the remaining two never outlining steps; and (3) running partial (RUN3) or complete (RUN4) solutions, which all but one of the AI group did for RUN3 and all for RUN4, versus 2 and 1, respectively, in the \NoAI group. Overall, even though some activities were delegated to AI, the use of AI did not dramatically eliminate the need for a broad range of detailed problem-solving behaviors to be engaged with by the participants.

Participants who delegated code writing or debugging responsibilities to AI (P3, P7, P11) had mixed results. For example, P7 was able to one-shot the \texttt{User.ts} file change required for one part of the overall task. However, P3 delegated the majority of their code writing to \AI and was not able to successfully complete the task. P11 was also able to delegate a small snippet of code in the same \texttt{User.ts} file as P7 to ChatGPT, although in the end they were unable to complete the full task. Interestingly, P2 and P10, who both successfully completed the task, did not delegate code generation to ChatGPT, but stuck with obtaining suggestions and coding the changes themselves with inspiration from the suggestions.

While the use of AI had some advantages, as discussed thus far in this section, it was not the only contributing factor to successful completion of the task. In analyzing the detailed activities of the ten participants, three additional contributing factors stood out.

\textit{Building a thorough understanding.} While everyone began the task by building an 
understanding of the system, P9 was the only participant who began with a top-down approach, beginning 
by testing the frontend's existing functionality before making changes, and tracing the flow of execution from the registration modal. This served them well because when they did not understand a TypeScript utility type in \texttt{internalClient.ts}, they had already traced through the execution and were able to deduce from that what data needed to be returned, so they could progress nonetheless.

\textit{Willingness to refocus.} P7's willingness to refocus while acting by switching to using \texttt{console.log} instead of remaining fixated trying to connect a debugger to VS Code was directly responsible for them finishing the task just before the allotted time expired. This is in comparison to P5, who continued ``messing around'' in the editor rather than searching or considering alternative strategies when they did not make the distinction that the frontend was running on a different port (3000) from the app's backend (3001).

\textit{Careful planning.} Careful planning led P10 to be highly successful in their use of \AI. Specifically, their concerted effort to break down and walk through the task methodically which, while leading to more \AI prompts than P3 (38 vs. 29), allowed them to keep better track of and integrate ChatGPT's suggestions. On the other hand, P3 nearly instantaneously began the task by providing instructions to \AI but with less explicit guidance, which led them and the \AI to become overwhelmed and confused by the problem's context.

\begin{finding}
{\textbf{Findings:}}
A majority of observed behaviors were performed by most of the AI and NoAI participants, with only a few activities that were unbalanced between the two groups. Partly this reflects the AI participants succeeded more often and thus were able to look back, and partly it reflects an offloading to AI instead of the Web (though not fully so). %
\end{finding}

\subsection{RQ2: Stuck Moments and Tools}
\label{rq2}

\begin{table}
\small
\centering
    \begin{tabular}{p{0.45\linewidth}p{0.46\linewidth}}
    \toprule
       \textbf{Stuck Type} & \textbf{Unstuck Type} \\
       \midrule
       \multicolumn{2}{l}{\textbf{Understanding}} \\
        \momentBox{momentLtYellow}{momentOrange}{S1:} Minimal Understanding\par
                        \momentBox{momentLtYellow}{momentOrange}{S2:} Flawed Understanding \par
                        \momentBox{momentLtGray}{momentGray}{TS1:} Flawed Understanding
                        &  \momentBox{momentLtBlue}{momentBlue}{US1:} Improve Understanding \par
                        \momentBox{momentLtBlue}{momentBlue}{US2:} Delegate to AI \par
                        \momentBox{momentLtGray}{momentGray}{TUS1:} AI overcomes flawed human understanding
                        \\
                        \midrule
        \multicolumn{2}{l}{\textbf{Making a plan}} \\                
        \momentBox{momentLtYellow}{momentOrange}{S3:} Flawed Plan\par
                        \momentBox{momentLtYellow}{momentOrange}{S4:} Know what to do but not how \par
                        \momentBox{momentLtGray}{momentGray}{TS2:} Suggests flawed plan\par
                        & \momentBox{momentLtBlue}{momentBlue}{US3:} Seek advice \par
                        \momentBox{momentLtBlue}{momentBlue}{US4:} Skip ahead \par
                        \momentBox{momentLtBlue}{momentBlue}{US5:} Trying alternatives \\
                        \midrule
    \multicolumn{2}{l}{\textbf{Acting on a plan}} \\
       \momentBox{momentLtYellow}{momentOrange}{S5:} Flawed execution with defective code & 
       \momentBox{momentLtGray}{momentGray}{TUS2:} Provides code to fix bug 
       \\
       \midrule
    \multicolumn{2}{l}{\textbf{Looking back}} \\
    \momentBox{momentLtYellow}{momentOrange}{S6:} Failure to look back \par
    \momentBox{momentLtYellow}{momentOrange}{S7:} Puzzled when looking back
        & \momentBox{momentLtBlue}{momentBlue}{US6:} Revisiting code or approach\\
       \bottomrule
    \end{tabular}
    \caption{Typology of \momentBox{momentLtYellow}{momentOrange}{stuck (S)} and \momentBox{momentLtBlue}{momentBlue}{unstuck (US)} moments in relation to Polya phases. The influence of \momentBox{momentLtGray}{momentGray}{tools} on getting stuck \momentBox{momentLtGray}{momentGray}{(TS)} or unstuck \momentBox{momentLtGray}{momentGray}{(TUS)} is also shown.}
    \label{tab:polya-stuck}
\end{table}

In answering our second research question, ``\textit{Why do developers get stuck and how can tools help or hinder them when they are stuck while making changes to unfamiliar code?}'', we 
firstly note that almost all participants got stuck at least once during the task execution (see ~\autoref{fig:polya_state_timeline}), with P9 as the sole exception.

\autoref{tab:polya-stuck} details different types of `stuck moments' occurring across Polya's phases and the approaches taken to becoming unstuck. In several cases, AI caused them to get stuck in the first place, exacerbated their stuckness, or helped them get unstuck. Other times---by exploring the code alongside the AI-generated advice---the developers avoided getting stuck. We now present each of these moments and, in so doing, note the involvement of AI. %
(The majority of our examples draw upon ChatGPT, but we note that web searches exhibited similar patterns of helping and hindering).

\paragraph{\momentBox{momentLtYellow}{momentOrange}{S1: Minimal understanding}}
This type of stuck moment involved the participant having little understanding of the application's architecture or code, resulting in an inability to make a plan.
An example is provided by P1 (NoAI), who begins the session by jumping into the first subtask: using the \texttt{createUser} function to persist the user registering into the database. After briefly examining the database initialization file, they repeatedly state over the next fifteen minutes that they are looking for examples of \texttt{createUser} (and frequently search the web for keywords such as ``js createUser'' and ``ts createUser''). 
Eventually, at the forty-minute mark, the participant gives up the search for functions by the same name and tries to determine which parameters their function should accept, and quickly finds an example online of how to insert a row into the database, finally becoming unstuck (\momentBox{momentLtBlue}{momentBlue}{US1}).

We can compare this example to P3 (AI), who, once they realized they were not advancing their understanding by perusing the codebase \momentBox{momentLtYellow}{momentOrange}{S2},  delegated code authoring to ChatGPT (\momentBox{momentLtBlue}{momentBlue}{US2}). They identified the code files to be changed, pasted them into ChatGPT, and asked ChatGPT to complete the function. In so doing, they made good progress and completed all the sub-tasks. AI, in a way, quietly filled the gaps in P3's understanding of the code.

\paragraph{\momentBox{momentLtYellow}{momentOrange}{S2: Flawed Understanding}} This kind of stuck moment occurs when the participant is unaware that their understanding of the task, code, or application is flawed. We refer to P3 (AI), who, in interacting with the AI to help them implement code, 
developed a flawed understanding of how the front-end web page interacted with the other code files. This caused problems when they tried to debug the application, as errors appeared in testing once all code had been written. Moreover, they reinforced the issue when they provided their flawed understanding in their prompts to ChatGPT, which led it to develop a similarly flawed understanding, making its suggestions unhelpful (\momentBox{momentLtGray}{momentGray}{TS1}). Ultimately, P3 was unable to complete the task, remaining stuck for a significant period.

Sometimes, however, AI helped participants even when they had a flawed understanding (\momentBox{momentLtGray}{momentGray}{TUS1}). P10 (AI) requested assistance with a code change, and ChatGPT correctly advised them on the change. P10 misunderstood the advice and attempted to change a different function, resulting in a semantic error message. They were stuck and asked ChatGPT for help, and ChatGPT continued to advise on the code change it initially suggested, stating \say{\textit{Then this is 100\% because ...}} (telling P10 they did it wrong).
After more back-and-forth---with ChatGPT patiently maintaining its position---P10 eventually understood and made the proper change. 

Other times, AI did not help overcome a participant's flawed understanding, as it also had a flawed understanding (\momentBox{momentLtGray}{momentGray}{TS1}). Consider P11 (AI), who was
struggling to determine how to link two functions necessary to complete a subtask. 
P11 started adding log statements to assess the application's flow, but then gave up and decided to delegate the work to ChatGPT.
ChatGPT incorrectly identified the snippet it was given by P11 as a \say{\textit{service-level or database-layer function}}, and recommended calling it directly.
P11 made the recommended change and considered their subtask complete, so they moved on. Later, during testing, an error
occurred because the front-end code was calling the controller directly. ChatGPT's incorrect assumption left P11 stuck for the next hour trying to debug the issue.

\paragraph{\momentBox{momentLtYellow}{momentOrange}{S3: Flawed plan}}
This type of stuck moment covers situations in which participants have a mostly correct understanding, but the approach they plan is invalidated by missing information or unknown constraints. For instance,
P10 (AI), prior to their stuck moment described earlier, %
had made a plan for how to add the registered user to the database. In thinking about how to do so, they asked ChatGPT not just for suggested code changes, but also, in a separate prompt, for a detailed explanation and rationale for those changes. Satisfied with the response, they spent a little bit more time figuring out what to do and then set out to do so. Unfortunately, P10 found a flaw in the plan co-designed with ChatGPT when VS Code highlighted some of the code they had written as an issue. The issue revealed that the earlier reasoning and thus plan were wrong.
Despite ChatGPT having provided them with a flawed plan, P10 nonetheless turned to ChatGPT to help them understand what went wrong (\momentBox{momentLtBlue}{momentBlue}{US3}). 
ChatGPT provided P10 with a ``fresh plan'': an alternatively structured solution that overcame the issue.

P10 also provides an example where the AI hindered the creation of a plan (\momentBox{momentLtGray}{momentGray}{TS2}), nearly leaving them further stuck. Because they did not provide ChatGPT with all defined interfaces, ChatGPT suggested modifying an existing interface mentioned in the prompt (erroneously). Fortunately, P10 did not make the change immediately. Instead, they dug into the code and used ChatGPT to assist, including providing it with the full set of defined interfaces. ChatGPT used this additional information to eventually propose the correct solution (to use a different interface). In this case, P10 properly disregarded ChatGPT's initial plan for what to change.

\paragraph{\momentBox{momentLtYellow}{momentOrange}{S4: Know what to do but not how}} The participant knows what they want to achieve, but does not have the knowledge necessary to %
complete the plan. P6 (NoAI) exemplifies this case of stuckness. They had correctly identified the line of code to change to insert a user into the database, but they had \say{\textit{forgotten how to insert a user}}. After 30 minutes of web searching, they were unable to determine the correct syntax for the insert statement. They decided to move on to the next subtask, thus skipping ahead (\momentBox{momentLtBlue}{momentBlue}{US4}).

P7 (AI), after having followed ChatGPT's incorrect plan of action, became stuck knowing they needed to debug their application, but not knowing how to do so. They sought advice by asking \AI, ``\textit{I am using yarn run dev to deploy the code, how to run it in debug mode}''. 
ChatGPT, however, gave P7 generic instructions, not specific to their situation. P7 attempted to follow these instructions, including clarifying their situation with \AI, but encountered environment issues when trying to set breakpoints. They continued debugging these for the following half hour. 
Eventually, P7 stops following the advice and resorts to basic log statements to progress (\momentBox{momentLtBlue}{momentBlue}{US5}).

\paragraph{\momentBox{momentLtYellow}{momentOrange}{S5: Flawed execution with defective code}}
This type of stuck moment is typified by participants who had developed
a good plan, but made an error in its execution. For example, P2
had a clear plan, a good understanding of the application, 
and asked very specific questions of ChatGPT that focused on TypeScript features, such as \texttt{Omit} in interfaces, and on unfamiliar components, such as Axios. P2 completed all the subtasks relatively quickly, only to find that the application threw an error when they tried to register a user. The root cause of the error was a buggy database call.
AI, however, was able to easily help P2 with the situation (\momentBox{momentLtGray}{momentGray}{TUS2}). P2 asked ChatGPT for assistance by providing the error code \say{\textit{Error: SQLITE\_CONSTRAINT: UNIQUE constraint failed: users.email}}. ChatGPT correctly diagnosed the issue as the user already existed in the database, \say{\textit{...because you are trying to insert a record with an email that already exists in the database}}. P2 initially does not believe ChatGPT, indicating cognitive dissonance between their understanding and ChatGPT's recommendation. After adding some \texttt{console.log} statements to debug, they realized ChatGPT was correct, updated the code, and successfully completed the task.

The situation of P11 (previously described) provides an interesting counterpoint, highlighting how AI can also hinder in this scenario due to poor advice stemming from the developer's flawed understanding (\momentBox{momentLtYellow}{momentOrange}{S2}). Part of the reason P11 remained stuck for an hour was that, once they were stuck, they tried to ask ChatGPT for help in numerous ways, including providing a code snippet, asking whether there was an error, and giving ChatGPT the specific error. ChatGPT dutifully answered each case. However, because P11 was unable to step outside their flawed understanding of the issue, the series of prompts to ChatGPT did not help P11 become unstuck.

\paragraph{\momentBox{momentLtYellow}{momentOrange}{S6: Failure to look back}}

In contrast to P11, who was stuck for a long time but at least attempted to step back, consider alternatives, and seek help in understanding what was wrong, P5 did not pause and reflect; in other words, they did not look back. They failed to question their understanding of the application's architecture, leading to persistent stuckness.

While P11 looked back and remained stuck, P10 became unstuck when they looked back. On observing syntax errors highlighted by VS Code, they tried various changes before asking ChatGPT to \say{\textit{Show me the last few code recommendations in the context of my ....}}, thereby reminding themselves of the various approaches. This appeared to jog their memory and they reverted the code, removing the syntax errors and becoming unstuck (\momentBox{momentLtBlue}{momentBlue}{US6}). 

\paragraph{\momentBox{momentLtYellow}{momentOrange}{S7: Puzzled when looking back}}
P2 ran the application after completing what they considered to be all necessary code changes. They kept getting an error message on the registration webpage and spent some time puzzled about what could be causing it. They remained befuddled for a little while, poking around at the code base, running it several times while poking some more, until eventually they realized that an identifier was not being returned and were able to complete the task.

\begin{finding}
{\textbf{Findings:}
Getting stuck can occur across all of Polya's phases. \AI can help become unstuck, but human insight is generally the driver for becoming unstuck. Unhelpfully, \AI can equally be the cause of being stuck when its advice is wrong or misinterpreted, and also can be the cause of remaining stuck when a shared flawed understanding persists in the human-AI interactions.}
\end{finding}

\section{Discussion of Research Contributions}
\label{discussion}

Our study makes three primary contributions to the literature on code understanding and the use of AI.
First, it shows how Polya's four phases can be a useful lens for understanding how AI (and other) tools impact code comprehension in the context of a realistic change task. Applying Polya's framework revealed that, even with AI tools accelerating writing code, significant challenges remained for developers when understanding the codebase and figuring out how to make the change.
Unpacking these challenges across Polya's phases through 25 inductive codes further revealed that many nuanced problem-solving behaviors remain essential.
Our analysis of stuck moments is particularly poignant: resolving stuck moments often became moments of joint human-AI sensemaking, where humans reinterpret AI outputs, search for a shared understanding, and reconstruct their mental models toward a common problem representation~\cite{HAO2025101152, JIANG2024100078}. Future work can use the framework and codes to understand how further advances in AI and other tools may alter nuanced developer behaviors that go beyond coding velocity.

Second, we observed that meta-cognitive skills are essential to success. Tools will change, and AI tools are changing the landscape of programming quickly~\cite{sauvola2024future}. 
Nonetheless, our analysis of observed behaviors indicated that the developers we studied benefited from AI, but also relied on critical thinking to succeed. In other words, the AI helped, but building a thorough understanding, a willingness to refocus, and careful planning were equally important. How tools are used, then, is more important than the tools themselves.
Similarly, being stuck was prolonged when participants steadfast pursued the same issue without stepping back and reflecting --- especially so when ChatGPT contributed to them remaining on the wrong path.
In this context, while we observed a majority of time being spent in understanding and acting, we suggest developers will benefit by spending more time planning and looking back.

Third, we contribute a typology of ``stuck moments'', detailing seven different ways in which developers got stuck on their tasks. The typology covers all four of Polya's phases, pointing to the fact that stuck moments are part and parcel of the problem-solving process for non-trivial code changes.
We further augmented our typology with possible mechanisms to become unstuck, as well as how AI in some cases helped and in other cases hindered becoming unstuck. We do not claim that the typology we present is complete.  Rather, we view it as an important starting point for others to expand upon and leverage the identified stuck moments to target tool development, developer training, and developer awareness.

We do not view our work as final or conclusive. Although the small number of participants allowed us to deeply investigate individual problem-solving behaviors, our findings must be interpreted in the context of our limited sample size as potential phenomena that must be examined in-depth with further studies in more focused, controlled settings.  Understanding the factors leading to task success, the impact of more time spent up front building understanding and planning for change, and how often developers in professional settings may become stuck and remain stuck at the hands of AI tools represent essential research questions for future research. It appears that AI enables those with a medium level of system understanding to complete tasks that those without AI fail to. Further study is necessary to understand whether these findings hold as tools evolve, as well as the long-term implications of a reduced level of system understanding.

At the end of the day, in the context of problem-solving in programming, \AI is just a tool. That said, 
the design of modern GenAI development tools 
has recently started to provide cognitive support for explaining unfamiliar code (e.g., ~\cite{Nam2024LlmCodeUnderstanding}) and our study reinforces that tools must offer comprehension-oriented assistance (see Table~\ref{table-codebook}). 
Our typology, and analysis of stuck moments in particular, offer a rich set of possibilities for what kinds of tools to design and build. AI or non-AI, tools should improve a developer's mental model and problem-solving skills, such as by helping form hypotheses about their codebase, parse logs/errors more quickly and effectively, outline steps to follow to accomplish a task, identify moments when developers are stuck and why, and more.  Generic LLMs can probably do this to some extent (e.g.,~\cite{balfroid_onboarding_2024} for explaining stack traces), though we suspect that future, more specialized tools could make a more significant difference.

Finally, our study leads to recommendations for education.  We found students already relied on AI, as some of those in the \NoAI group commented they could not do the task without it, even though they could search on the internet.
But even with AI use, participants struggled to build accurate mental models of the system architecture and to debug their solutions. Our observations suggest that developers need guidance on how to use AI to support building useful models of the codebase, and strengthen their meta-cognitive awareness and reflective skills when progress is not going as planned. Junior developers in particular tend ``to fail to recognize when they are truly stuck and should ask for help''~\cite{begel2008b}, and need to be taught to look back more often.
As educators, we still have much to learn about how we can teach student developers to use these AI tools effectively so they develop genuine understanding rather than outsourcing their cognition to the tool.  This is particularly important in the context of ChatGPT giving the wrong answers, which our study shows participants did not necessarily spot right away and could be seriously detrimental to progress. 
This indicates that we must continue to teach how to build one's own understanding.

\section{Limitations and Threats to Validity}
\label{sec:limitations}

As a small, exploratory laboratory study, our findings should be interpreted alongside several threats to validity.

\textit{Construct Validity.}
Although we used Polya’s framework, and inductive codes, as a proxy for developers’ problem-solving and comprehension processes, we did not directly measure cognitive states or the accuracy of their understanding beyond think-aloud data, architectural sketches, and post-task responses. This limits the construct validity of our interpretations. However, given that this is the first study to use Polya's framework to capture developers' problem-solving processes, it also provides a valuable foundation for future studies. Future research could incorporate more direct cognitive measures (e.g., comprehension quizzes or cognitive load instruments) to triangulate and extend our findings.

We used ChatGPT as a representative GenAI tool. Other GenAI tools may produce different results, and the model version available during the study reflects a rapidly evolving technology landscape; this limitation applies broadly to all current GenAI research.

Participants recognized that their task was fictional and acted accordingly, e.g., multiple participants chose to forego hashing the password they were storing in the database, stating variations on ``I wouldn't normally do this''. Similarly, the artificial time constraint may have changed participants' tendency to test as they go. Such cutting corners may or may not impact problem-solving and future studies should consider how to account for it.

\textit{Internal Validity.}
Participant expertise varied across programming maturity and knowledge of the technologies used (TypeScript, React, Express). This variability added learning challenges beyond program comprehension. It thus complicates attribution of outcomes solely to AI support, though this was not our study's focus.

The detailed task instructions---including subtasks and linked references---ensured feasibility, but may have constrained planning behavior relative to less structured environments. 
Participants whose sole motivation was to finish the task may also have looked back less.
Although we employed iterative coder alignment procedures, qualitative interpretation of observational data remains subject to researcher bias, and observing cognitive behavior is challenging as much of what happens occurs in the developers' heads.

\textit{External Validity.}
Our participants were predominantly upper-year university students, who served as a reasonable but imperfect proxy for novice professional developers. Junior developers frequently encounter coding tasks on unfamiliar codebases similar to our scenario, but we do not claim full generalizability to experienced practitioners or other development contexts. We also do not claim full saturation of stuck and unstuck moments shown in the typology in~\autoref{tab:polya-stuck}. Our small sample size was a deliberate methodological choice enabling deep qualitative analysis, though it limits the breadth of observed behaviors.

\section{Conclusion and Future Work}
\label{conclusion}

By applying the lens of Polya's problem-solving process to examine how developers implemented a change in an unfamiliar code base, our study offers new insights into how GenAI tools such as ChatGPT can shape program comprehension and problem-solving strategies. While ChatGPT 
use leads to 
fewer code modifications, nearly all developers experienced frequent and varied forms of ``stuckness'', underscoring that AI assistance does not eliminate cognitive challenges in software development but rather reshapes them. Our typology of stuck moments further revealed that ChatGPT can act as a friend (e.g., suggest solutions, overcome gaps in an incomplete understanding) or foe (e.g., persist with incorrect advice, simply answer with what is asked for). This highlights opportunities for tools that better support global system understanding, hypothesis-driven debugging, and recovery from impasses. Moreover, it suggests educational practices need to incorporate problem-solving and effective AI-assisted debugging strategies, and that new tools should be considered to support developer comprehension and resilient problem-solving skills in the age of AI powered development.

\begin{acks}
Thank you to the study participants; Rena Kollmann-Suhr (initial visualizations); Nathan Cassee, Arty Starr, and Quinton Yong (suggested literature); and CHISEL lab (regular feedback and support).
This material is based on work supported by the National Science Foundation under grants CCF-2210812 and CCF-2210813; the Federal Institute of Education, Science and Technology of Rio Grande do Sul (IFRS) and by the Ministry of Science, Technology, and Innovation of Brazil (Law 8.248 from Oct 23, 1991), within the scope of PPI-SOFTEX, coordinated by Softex, and published in the Residência em TIC 02 - Aditivo, Official Gazette 01245.012095/2020-56; Natural Sciences and Engineering Research Council of Canada (NSERC), application IDs 594057-2024 and 606783/2025; and partially financed by the University of Zurich and Digital Society Initiative.
\end{acks}

\bibliographystyle{ACM-Reference-Format}
\bibliography{refs}

\end{document}